\documentclass[a4paper,11pt]{article}
\pdfoutput=1 
\usepackage{graphics}
\usepackage[utf8]{inputenc}
\usepackage{jheppub}
\usepackage{epsfig}
\usepackage{subfig}
\usepackage{float}
\usepackage{slashed}
\usepackage{soul}
\usepackage{cancel}
\usepackage{multirow}
\usepackage{amsmath,amssymb,mathrsfs}
\usepackage{color}
\definecolor{ourcolor}{rgb}{0.7, 0.25, 0.05}

\def\be{\begin{equation}}
\def\ee{\end{equation}}
\def\al{\alpha}
\def\bea{\begin{eqnarray}}
\def\eea{\end{eqnarray}}
\def\M{M_{\mbox{\tiny Pl}}}

\title{\color{ourcolor}{Lorentz invariance violation as an explanation of muon excess in Auger data}}

\author[]{Gaurav Tomar}
\affiliation[]{Physical Research Laboratory, Ahmedabad 380009, India.}

\emailAdd{tomar@prl.res.in}

\abstract{The Auger collaboration has observed the number of muons which is higher than its prediction by existing hadronic interaction models. We explain this excess of muons by using Lorentz invariance violation (LIV) in photon sector. As an outcome of Lorentz invariance violation, the dispersion relation of photon gets modified, which we use for the calculation of $\pi^0$ decay width. In the Auger data of primary energy $10^{9.8}<E(\mbox{GeV})<10^{10.2}$, we find that the neutral pion decay width is suppressed in comparison to its standard model (SM) counterpart. As a result, we get a large number of muons explaining the observed muon excess. We consider Planck suppressed LIV at order $\mathcal{O}(p^2/M^2_{\tiny \mbox{Pl}})$ for studying the photon sector, which is in agreement with the current bounds, and not as tightly constrained as LIV at order $\mathcal{O}(p/M_{\tiny \mbox{Pl}})$.
}

\allowdisplaybreaks

\begin{document}

\maketitle




\bigskip

\section{Introduction}
\label{sec:intro}
Ultra high energy (UHE) cosmic rays with energy of $\sim\mathcal{O}(10^{11})$ GeV are the most energetic particles observed on the Earth. After their collisions with the Earth atmosphere, a huge cascade of secondary particles with low energy is created. As the collision energy is roughly ten times higher than the one at LHC, it can be a suitable window for new physics. These cascades of particles or showers are explored by large arrays like Yakutsk Extensive Air Shower Array and Pierre Auger Observatory (PAO). A new study from Auger collaboration~\cite{Aab2015a, Aab2016} suggests that the number of muons produced in UHE showers is higher in comparison to the one predicted by existing  models~\cite{Abu-Zayyad2000, Aab2014, Aab2015}. Basically, the hadronic component of showers with primary energy $10^{9.8}<E(\mbox{GeV})<10^{10.2}$ have $30\%$-$60\%$ more muons than expected~\cite{Aab2015a, Aab2015b, Aab2016}.\\
The explanation of the muon excess in PAO data is challenged by the distribution of the depth of shower maximum, $X_{max}$, which should be independently fitted. To fit the data, the density of muon at 1 km from the shower core which is denoted as, $N_\mu$, should increase. The properties of hadronic interactions which affect $N_\mu$ and $X_{max}$ are: cross-section, elasticity, multiplicity, primary mass, and $\pi^0$ energy fraction. The variation of $N_\mu$ and $X_{max}$ with them is shown in~\cite{Allen2013}, where it is noted that changing $\pi^0$ energy fraction is the only viable option for increasing $N_\mu$. Any other change in $N_\mu$ without affecting the longitudinal profile $X_{max}$ is not possible which is tightly constrained (see fig.~1 of ref.~\cite{Allen2013}). If hadronic shower carries $f_{\mbox{\tiny had}}$ energy fraction of the total primary cosmic ray energy $E$, then it scales as,
\be
 f_{\mbox{\tiny had}} \sim (1-f_{\mbox{\tiny EM}})^{n_{\mbox{\tiny gen}}}, 
 \label{eq:ehfr}
\ee
where $f_{\mbox{\tiny EM}}$ is the fraction of energy transfered into electromagnetic particles per generation, and $n_{\mbox{\tiny gen}}$ is the number of generations required for most pions to have energy below $\sim 100$ GeV. Below 100 GeV energy, most of the charged pions decay rather than interact, terminating the energy transfer to the electromagnetic component of the shower. While the charged pions interact instead of decay above $100$ GeV energy, persisting the hadronic shower.
The best way to increase $f_{\mbox{\tiny had}}$ is to reduce either $n_{\mbox{\tiny gen}}$ or $f_{\mbox{\tiny EM}}$. The estimated value of $n_{\mbox{\tiny gen}}$ needed to reach pion energy below $\sim 100$ GeV is $n_{\mbox{\tiny gen}}=3,4,5,6$ for primary energy $E=10^5,10^6,10^7,10^8$ GeV respectively~\cite{Matthews2005}. But the required $n_{\mbox{\tiny gen}}$ for getting the desired result also reduce $X_{max}$ which is tightly constrained. So the best option for increasing the muon density is to reduce $f_{\mbox{\tiny EM}}$ ($\pi^0$ energy fraction) \footnote{In ref.~\cite{Diaz2016}, the variation of $X_{max}$ as a function of photon energy is discussed in LIV framework. We will examine this point in Sec.~\ref{sec:discuss}.}.\\
There are many proposals for reducing $\pi^0$ energy fraction such as, chiral symmetry restoration, pion decay suppression, and pion production suppression~\cite{Allen2013, Farrar2013}. The string percolation models~\cite{Alvarez-Muniz2012} and strange fireball mechanism~\cite{Anchordoqui2016} are other approaches used for the explanation of the observed muon excess. In this work, we focus on the decay suppression of $\pi^0$ which can occur from Lorentz invariance violation in photon sector. As the lifetime of $\pi^0$ is very small $\sim \mathcal{O}(10^{-17}~\mbox{sec})$, it decays immediately into two photons after its production. We modify the photon dispersion relation in the spirit of~~\cite{Galaverni2008, Maccione2008, Galaverni2008a}, and calculate the neutral pion decay width. 
At high energy,
as a result of modified dispersion relation, photon becomes massive enough to suppress the $\pi^0$ decay into two photons.
The possible Lorentz invariance violation is motivated from quantum gravity~\cite{Mattingly2005, Liberati2013, Cognola2016} and in many studies~\cite{Ellis2004, Horava:2009uw, Mohanty2011, Mohanty2012, Girelli2012, Anchordoqui2014a, Tomar2015} it has been shown that LIV becomes important at very high energy scale. There are stringent constraints on LIV in photon~\cite{Galaverni2008, Maccione2008, Galaverni2008a, Kostelecky2011} and fermion~\cite{Gagnon2004, Scully2009, Bi2009, Maccione2009}. Specifically, LIV in photon sector is tightly constrained for Planck mass suppressed dim-5 operators and even dim-6 operators are constrained to a unprecedented level~\cite{Galaverni2008, Maccione2008, Kostelecky2011}. For dim-6 operators, the bound on photon LIV parameter $\eta$ is, $\eta \gtrsim -10^{-7}$~\cite{Galaverni2008} which comes from the stringent upper bound on photon flux above $10^{11}$ GeV~\cite{Rubtsov2006}, and if photon is observed at $10^{10}$ GeV then $\eta \lesssim 10^{-8}$~\cite{Maccione2008}. We consider a dim-6 scenario (LIV at order $\mathcal{O}(p^2/M^2_{\tiny \mbox{Pl}})$) in this work and find that for getting the desired muon excess, LIV parameter is $\eta \sim 10^{-2}$, which seems to be in tension with the upper bound mentioned in~\cite{Maccione2008, Galaverni2008a}. But we want to emphasize that the upper limit quoted in~\cite{Maccione2008, Galaverni2008a} is based on the assumption of the observation of photon with $10^{10}$ GeV energy. In cosmic rays, photon with this much energy is a question of discussion~\cite{Abbasi2016, Aab2016a}, and the upper bound can be avoided at present. 
The rest of the paper is as follows: in Sec.~\ref{sec:dr}, we discuss the modified dispersion relation. We give the neutral pion decay calculation in LIV framework in Sec.~\ref{sec:pidky}, and our discussion and conclusion in Sec.~\ref{sec:discuss} and Sec.~\ref{sec:concl} respectively.
\section{Dispersion relation}
\label{sec:dr}
The Lorentz invariance violation modifies the dispersion relation of photon. As we mentioned before, LIV at order $\mathcal{O}(p/M_{\tiny \mbox{Pl}})$ corresponds to a cubic dispersion relation which arises from dim-5 operator. The LIV at order $\mathcal{O}(p/M_{\tiny \mbox{Pl}})$ is tightly constrained with the required suppression scale well above the Planck mass. In the following, we consider the underlying theory to be $CPT$ invariant by taking LIV at order $\mathcal{O}(p^2/M^2_{\tiny \mbox{Pl}})$.
We denote the 4-momentum of the photon $\gamma(p_1)$ by $(E_1,\vec{p_1})$ and consider the following dispersion relation for photon,
\be
 E^2_1 = p^2_1 + \eta p^2_1 \left(\frac{p_1}{\M}\right)^n,
 \label{eq:dr}
\ee
where $\eta$ is a LIV parameter and Planck mass $\M=1.2 \times 10^{19}$ GeV. This dispersion relation can be obtained from the Lagrangian given in~\cite{Mattingly2008}. As $n=1$ scenario of eq.~(\ref{eq:dr}) arises from $CPT$-odd contributions~\cite{Jacobson2003, Myers2003, Jacobson2006}, it is tightly constrained~\cite{Maccione2007, Galaverni2008, Kostelecky2011}. In the following, we assume that theory is $CPT$-even by taking $n=2$.
\section{Neutral pion decay}
\label{sec:pidky}
\begin{figure}[!htbp]
\begin{center}
\includegraphics[angle=360,scale=0.75]{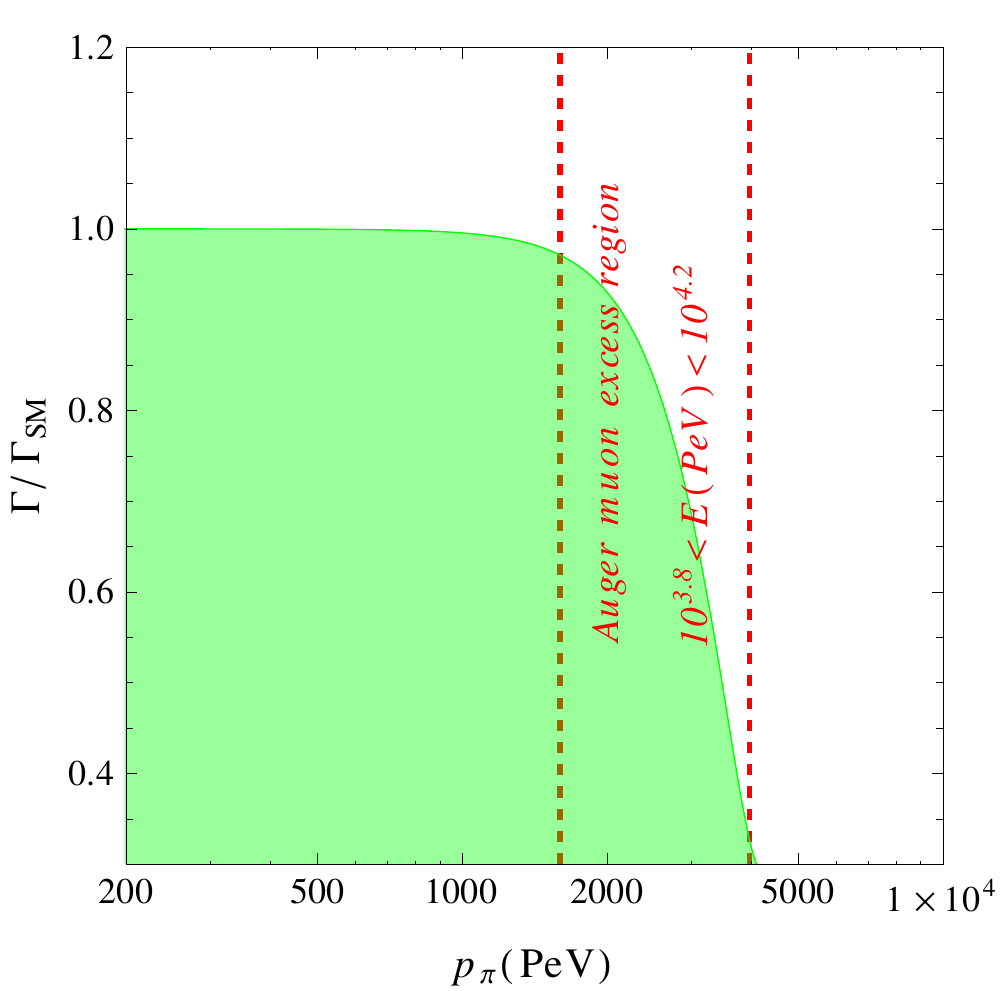}
   \caption{The ratio $\Gamma/\Gamma_{\mbox{\tiny SM}}$ for $\pi^0\rightarrow \gamma \gamma$ process in Lorentz invariance violating framework to its SM counterpart as a function of pion momentum $p_\pi$ by considering $\eta=10^{-2}$. The Auger region where the muon excess is observed also shown.}
\label{fig:dwc}
\end{center}
\end{figure}

We calculate the neutral pion decay width using modified dispersion relation of eq.~(\ref{eq:dr}) considering $n=2$. We compute the amplitude for neutral pion decay process $\pi^0(q)\rightarrow \gamma(p_1) \gamma(p_2)$, which is dominated by chiral anomaly and reads~\cite{Bernstein2013},
\be
 \mathcal{M} = \frac{e^2}{4\pi^2 f_\pi}\epsilon_{\mu\nu\alpha\beta}\epsilon^\mu_1\epsilon^\nu_2 p^\alpha_1 p^\beta_2,
\ee
where $f_\pi$ is the pion decay constant. We calculate the average amplitude square, which is,
\be
 |\mathcal{M}|^2 = \frac{e^4}{64\pi^4 f^2_\pi} (m^2_{\pi}-\eta^\prime p^4_1-\eta^\prime p^4_2)^2,
\ee
where $\eta^\prime \equiv \eta/\M^2$. The decay width of $\pi^0$ is then given as,
\begin{align}
 \Gamma =& \frac{\al^2}{64\pi^3 f^2_\pi E_\pi}\int \frac{p_1 dp_1 d\rm cos\theta}{\sqrt{|\vec{p}-\vec{p_1}|^2+\eta^\prime |\vec{p}-\vec{p_1}|^4}}
          \delta(E_\pi-E_1-\sqrt{|\vec{p}-\vec{p_1}|^2+\eta^\prime |\vec{p}-\vec{p_1}|^4})\nonumber\\
          &\times (m^2_{\pi}-\eta^\prime p^4_1-\eta^\prime (p-p_1)^4)^2,
\label{eq:dw1}          
\end{align}
where $\alpha=e^2/4\pi$ is the fine structure constant and $E_1$ is the photon energy which is defined as $E_1=\sqrt{p^2_1+\eta^\prime p^4_1}$. The momentum of photon is defined as, $|\vec{p}-\vec{p_1}|^2=p^2+p^2_1-2p p_1~\rm cos\theta$. From the argument of delta function in eq.~(\ref{eq:dw1}), one reads,
\be
 \sqrt{|\vec{p}-\vec{p_1}|^2+\eta^\prime |\vec{p}-\vec{p_1}|^4} = E_\pi-E_1,
\ee
which after solving gives,
\be
 {\rm cos\theta} = \frac{2p_1 E_\pi-m^2_\pi + \eta^\prime(E^4_\pi-4E^3_\pi p_1 + 6E^2_\pi p^2_1-3E_\pi p^3_1)}{2p p_1}.
 \label{eq:cost}
\ee
We reduce the argument of $\delta$ function in terms of $\rm cos\theta$ by taking,
\be
 \left|\frac{d}{d\rm cos\theta}(E_\pi-E_1-\sqrt{|\vec{p}-\vec{p_1}|^2+\eta^\prime |\vec{p}-\vec{p_1}|^4})\right|=\frac{p p_1}{\sqrt{p^2 + p^2_1-2p p_1 {\rm cos\theta}+\eta^\prime |\vec{p}-\vec{p_1}|^4}}.
\ee
After these simplifications, we get the decay width of neutral pion,
\begin{align}
  \Gamma =& \frac{\al^2}{64\pi^3 f^2_\pi E_\pi}\int \frac{dp_1}{p} (m^2_{\pi}-\eta^\prime p^4_1-\eta^\prime \tilde p^4_2)^2,
  \label{eq:dw2}
\end{align}
where $\tilde p^4_2 = m^4_\pi+2 m^2_\pi p^2 + p^4-4 E_\pi m^2_\pi p_1-4 E_\pi p^2 p_1+4 E^2_\pi p^2_1+2 m^2_\pi p^2_1+2 p^2 p^2_1-4 E_\pi p^3_1 + p^4_1$. We perform the integration of eq.~(\ref{eq:dw2}) in the allowed limits of $p_1$, which are fixed by taking $\rm cos~\theta=\pm 1$ in eq.~(\ref{eq:cost}), and gives,
\be
 p_{1_{\rm max}}=\frac{m^2_\pi-\eta^\prime(E^4_\pi-4 E^3_\pi p_{1_{\rm max}}+6 E^2_\pi  p^2_{1_{\rm max}}-3E_\pi p^3_{1_{\rm max}})}{2(E_\pi-p)},
\ee
\be
 p_{1_{\rm min}}=\frac{m^2_\pi-\eta^\prime(E^4_\pi-4 E^3_\pi p_{1_{\rm min}}+6 E^2_\pi  p^2_{1_{\rm min}}-3E_\pi p^3_{1_{\rm min}})}{2(E_\pi + p)}.
\ee
By solving these equations numerically, we get the allowed limits on the photon momentum. Using these limits, we solve eq.~(\ref{eq:dw2}) to get the decay width of $\pi^0$ and then compare it with the SM result of pion decay in a moving frame, which is given as,
\be
 \Gamma_{\mbox{\tiny SM}}(\pi^0 \rightarrow \gamma \gamma) = \frac{\al^2 m^4_\pi}{64\pi^3 f^2_\pi E_\pi}.
 \label{eq:dwsm}
\ee
In fig.~(\ref{fig:dwc}), we have shown the deviation of $\pi^0$ decay width from its SM prediction (see eq.~(\ref{eq:dwsm})). We find that as a result of phase space and $|\mathcal{M}|^2$ suppression, the decay width of $\pi^0$ (electromagnetic energy transfered per generation, $f_{\mbox{\tiny EM}}$) decreases with large pion momentum. As a result, $f_{\mbox{\tiny had}}$ increases (see eq.~(\ref{eq:ehfr})), which can enhance the number of muons by $30\%-60\%$ in the desired energy range. We have shown the Auger muon excess region for the primary cosmic ray energy $E$ $(10^{3.8}<E~(\mbox{PeV})<10^{4.2})$, which translate into neutral pion energy with $\sim 25\% E$~\cite{GarciaCanal2009}.
 Here it is important to mention that we also checked our calculation for $n=1$ scenario and found that $\eta \sim 10^{-12}$ is required to explain the Auger muon excess, which is three orders of magnitude higher than the current bounds~\cite{Kostelecky2011}. So it is not possible to address the observed muon excess in Auger data for $n=1$ scenario.
\section{Discussion}
\label{sec:discuss}
\begin{figure}[!htbp]
\begin{center}
\includegraphics[angle=360,scale=1.0]{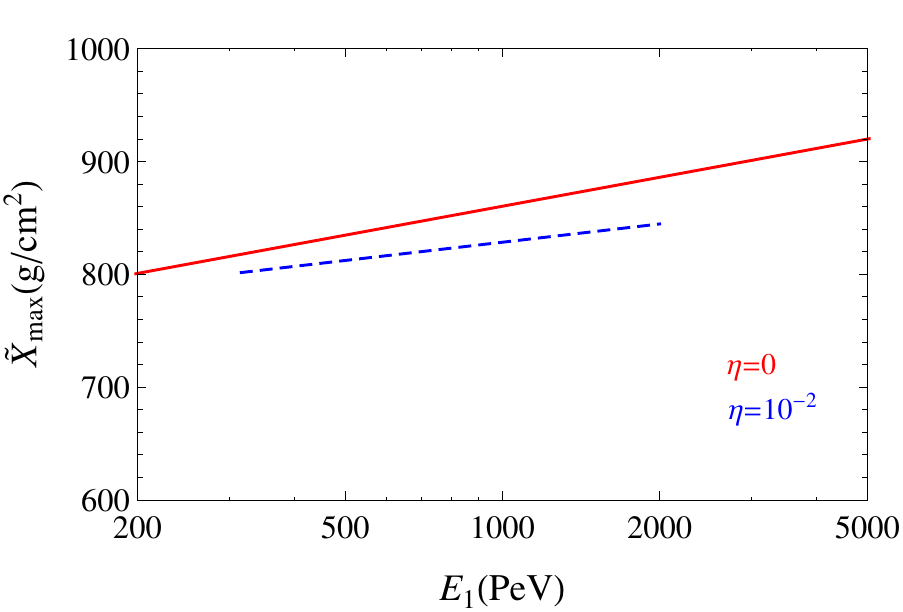}
   \caption{The electromagnetic shower depth $\tilde X_{max}$ as a function of initial photon energy $E_1$ for LIV parameter $\eta=10^{-2}$. The Lorentz invariant standard result $(\eta=0)$ is also shown for comparison.}
\label{fig:shmax}
\end{center}
\end{figure}

 In the previous sections, we discussed how modified dispersion relation gives rise to a massive photon, which stops $\pi^0$ decay at energy denoted as $E^{\rm cutoff}_\pi$. We mentioned that $\pi^0$ decay does not contribute into shower maximum depth $X_{max}$, but as a result of LIV, it is possible that photon becomes massive enough to decay into $e^+e^-$ pairs which can modify $X_{max}$.
 We contemplate this idea in the spirit of~\cite{Diaz2016}, and check our LIV scenario against that. The threshold energy for photon decay into $e^+e^-$ pairs is,
 \be
  E_{1_{\rm th}} \simeq \sqrt{\frac{2 m_e \M}{\eta^{1/2}}},
  \label{eq:photh}
 \ee
 where $m_e$ is the mass of electron. We get $E_{1_{\rm th}}$ after considering the condition $m_\gamma \simeq 2 m_e$. If the initial photon energy $E_{1} >  E_{1_{\rm th}}$, then photon starts decaying into pair of $e^+e^-$. As a result, the shower maximum gets modified and can be written as~\cite{Diaz2016},
 \be
  \tilde X_{max} = \lambda_r \beta~{\rm ln}\left(\frac{E_1/2}{ E_{1_{\rm th}}}\right) + \lambda_r~{\rm ln}\left(\frac{ E_{1_{\rm th}}}{E_c}\right),
 \ee
 where $E_c$ is the critical energy at which ionization starts dominating over radiative processes, $\lambda_r$ is the radiation length in the medium, and $\beta\equiv \rm ln2/ln3$. In the standard scenario ($\eta=0$), $X_{max}=\lambda_r~{\rm ln}(E_1/E_c)$~\cite{Letessier-Selvon2011}. In fig.~(\ref{fig:shmax}), we have shown the behavior of modified $\tilde X_{max}$ as a function of initial photon energy $E_1$ by using $E_c\approx 80$ MeV, $\lambda_r \approx 37~\rm g/cm^3$~\cite{Letessier-Selvon2011}, and $\eta=10^{-2}$. The standard Lorentz invariant $(\eta=0)$ scenario is also shown for comparison. By taking $\eta=10^{-2}$, eq.~(\ref{eq:photh}) gives $ E_{1_{\rm th}}\approx 315$ PeV. It is clear from fig.~(\ref{fig:dwc}) that neutral pion decay for $\eta= 10^{-2}$ stops at $E^{{\rm cutoff}}_\pi\sim 4000$ PeV, so there should not be any photon with energy $E_1 > E^{{\rm cutoff}}_\pi/2$. Analyzing fig.~(\ref{fig:shmax}), we find that in the allowed region $E_{1_{\rm th}}  < E_1 < E^{{\rm cutoff}}_\pi/2$, the modified shower maximum depth $\tilde X_{max}$ varies between 10-40$~\rm g/cm^3$. The precise measurement of $X_{max}$ in future can be used to probe this model. 
\section{Conclusion}
\label{sec:concl}
In this paper, we discuss the Lorentz invariance violation explaining the muon excess observed by Auger collaboration. The relative number of muons can be increased either by reducing the energy fraction in electromagnetic decay i.e. suppressing the neutral pion decay or reducing the $n_{\mbox{\tiny gen}}$. As the variation of $n_{\mbox{\tiny gen}}$ is tightly constrained from the independent observation of $X_{max}$, change in $f_{\mbox{\tiny EM}}$ is the best option for increasing the number of muons. We reduce the energy fraction $f_{\mbox{\tiny EM}}$ by suppressing the $\pi^0$ decay width. We consider the modified dispersion relation for photon by taking LIV at order $\mathcal{O}(p^2/M^2_{\tiny \mbox{Pl}})$ from a $CPT$-even dim-6 operator, and calculate the neutral pion decay width in this scenario. We find that, at high energies, Lorentz invariance violation starts playing an important role and suppress the decay width of $\pi^0$, which depends on the value of LIV parameter $\eta$. We find that by taking Plank mass square suppressed $\eta\sim 10^{-2}$, it is possible to suppress the decay width of $\pi^0$ in the desired energy range. As a result of neutral pion decay width suppression, the energy fraction in electromagnetic shower reduces and it gives rise to the relative number of muons observed by Auger collaboration. 
\section{Acknowledgement}
The  author would like to thank Subhendra Mohanty, Namit Mahajan and Petr Satunin for helpful discussions. The author also thanks Ujjal Kumar Dey for reading the manuscript and anonymous referee for his/her constructive comments on the manuscript.
\bibliographystyle{JHEP}
\bibliography{TomarRef}

\providecommand{\href}[2]{#2}\begingroup\raggedright\begin{thebibliography}{10}

\bibitem{Aab2015a}
{\bf Pierre Auger} Collaboration, A.~Aab et~al., {\it {The Pierre Auger Cosmic
  Ray Observatory}},  {\em Nucl. Instrum. Meth.} {\bf A798} (2015) 172--213,
  [\href{http://arxiv.org/abs/1502.01323}{{\tt arXiv:1502.01323}}].

\bibitem{Aab2016}
{\bf Pierre Auger} Collaboration, A.~Aab et~al., {\it {Testing Hadronic
  Interactions at Ultrahigh Energies with Air Showers Measured by the Pierre
  Auger Observatory}},  {\em Phys. Rev. Lett.} {\bf 117} (2016), no.~19 192001,
  [\href{http://arxiv.org/abs/1610.08509}{{\tt arXiv:1610.08509}}].

\bibitem{Abu-Zayyad2000}
{\bf MIA, HiRes} Collaboration, T.~Abu-Zayyad et~al., {\it {Evidence for
  Changing of Cosmic Ray Composition between 10**17-eV and 10**18-eV from
  Multicomponent Measurements}},  {\em Phys. Rev. Lett.} {\bf 84} (2000)
  4276--4279, [\href{http://arxiv.org/abs/astro-ph/9911144}{{\tt
  astro-ph/9911144}}].

\bibitem{Aab2014}
{\bf Pierre Auger} Collaboration, A.~Aab et~al., {\it {Depth of maximum of
  air-shower profiles at the Pierre Auger Observatory. II. Composition
  implications}},  {\em Phys. Rev.} {\bf D90} (2014), no.~12 122006,
  [\href{http://arxiv.org/abs/1409.5083}{{\tt arXiv:1409.5083}}].

\bibitem{Aab2015}
{\bf Pierre Auger} Collaboration, A.~Aab et~al., {\it {Muons in air showers at
  the Pierre Auger Observatory: Mean number in highly inclined events}},  {\em
  Phys. Rev.} {\bf D91} (2015), no.~3 032003,
  [\href{http://arxiv.org/abs/1408.1421}{{\tt arXiv:1408.1421}}]. [Erratum:
  Phys. Rev.D91,no.5,059901(2015)].

\bibitem{Aab2015b}
{\bf Pierre Auger} Collaboration, A.~Aab et~al., {\it {Muons in air showers at
  the Pierre Auger Observatory: Mean number in highly inclined events}},  {\em
  Phys. Rev.} {\bf D91} (2015), no.~3 032003,
  [\href{http://arxiv.org/abs/1408.1421}{{\tt arXiv:1408.1421}}]. [Erratum:
  Phys. Rev.D91,no.5,059901(2015)].

\bibitem{Allen2013}
J.~Allen and G.~Farrar, {\it {Testing models of new physics with UHE air shower
  observations}},  in {\em {Proceedings, 33rd International Cosmic Ray
  Conference (ICRC2013): Rio de Janeiro, Brazil, July 2-9, 2013}}, p.~1182,
  2013.
\newblock \href{http://arxiv.org/abs/1307.7131}{{\tt arXiv:1307.7131}}.

\bibitem{Matthews2005}
J.~Matthews, {\it {A Heitler model of extensive air showers}},  {\em Astropart.
  Phys.} {\bf 22} (2005) 387--397.

\bibitem{Diaz2016}
J.~S. Diaz, F.~R. Klinkhamer, and M.~Risse, {\it {Changes in extensive air
  showers from isotropic Lorentz violation in the photon sector}},  {\em Phys.
  Rev.} {\bf D94} (2016), no.~8 085025,
  [\href{http://arxiv.org/abs/1607.02099}{{\tt arXiv:1607.02099}}].

\bibitem{Farrar2013}
G.~R. Farrar and J.~D. Allen, {\it {A new physical phenomenon in ultra-high
  energy collisions}},  {\em EPJ Web Conf.} {\bf 53} (2013) 07007,
  [\href{http://arxiv.org/abs/1307.2322}{{\tt arXiv:1307.2322}}].

\bibitem{Alvarez-Muniz2012}
J.~Alvarez-Muniz, L.~Cazon, R.~Conceicao, J.~D. de~Deus, C.~Pajares, and
  M.~Pimenta, {\it {Muon production and string percolation effects in cosmic
  rays at the highest energies}},  \href{http://arxiv.org/abs/1209.6474}{{\tt
  arXiv:1209.6474}}.

\bibitem{Anchordoqui2016}
L.~A. Anchordoqui, H.~Goldberg, and T.~J. Weiler, {\it {Strange fireball as an
  explanation of the muon excess in Auger data}},
  \href{http://arxiv.org/abs/1612.07328}{{\tt arXiv:1612.07328}}.

\bibitem{Galaverni2008}
M.~Galaverni and G.~Sigl, {\it {Lorentz Violation in the Photon Sector and
  Ultra-High Energy Cosmic Rays}},  {\em Phys. Rev. Lett.} {\bf 100} (2008)
  021102, [\href{http://arxiv.org/abs/0708.1737}{{\tt arXiv:0708.1737}}].

\bibitem{Maccione2008}
L.~Maccione and S.~Liberati, {\it {GZK photon constraints on Planck scale
  Lorentz violation in QED}},  {\em JCAP} {\bf 0808} (2008) 027,
  [\href{http://arxiv.org/abs/0805.2548}{{\tt arXiv:0805.2548}}].

\bibitem{Galaverni2008a}
M.~Galaverni and G.~Sigl, {\it {Lorentz Violation and Ultrahigh-Energy
  Photons}},  {\em Phys. Rev.} {\bf D78} (2008) 063003,
  [\href{http://arxiv.org/abs/0807.1210}{{\tt arXiv:0807.1210}}].

\bibitem{Mattingly2005}
D.~Mattingly, {\it {Modern tests of Lorentz invariance}},  {\em Living Rev.
  Rel.} {\bf 8} (2005) 5, [\href{http://arxiv.org/abs/gr-qc/0502097}{{\tt
  gr-qc/0502097}}].

\bibitem{Liberati2013}
S.~Liberati, {\it {Tests of Lorentz invariance: a 2013 update}},  {\em Class.
  Quant. Grav.} {\bf 30} (2013) 133001,
  [\href{http://arxiv.org/abs/1304.5795}{{\tt arXiv:1304.5795}}].

\bibitem{Cognola2016}
G.~Cognola, R.~Myrzakulov, L.~Sebastiani, S.~Vagnozzi, and S.~Zerbini, {\it
  {Covariant Horava-like and mimetic Horndeski gravity: cosmological solutions
  and perturbations}},  {\em Class. Quant. Grav.} {\bf 33} (2016), no.~22
  225014, [\href{http://arxiv.org/abs/1601.00102}{{\tt arXiv:1601.00102}}].

\bibitem{Ellis2004}
J.~R. Ellis, N.~E. Mavromatos, D.~V. Nanopoulos, and A.~S. Sakharov, {\it
  {Space-time foam may violate the principle of equivalence}},  {\em Int. J.
  Mod. Phys.} {\bf A19} (2004) 4413--4430,
  [\href{http://arxiv.org/abs/gr-qc/0312044}{{\tt gr-qc/0312044}}].

\bibitem{Horava:2009uw}
P.~Horava, {\it {Quantum Gravity at a Lifshitz Point}},  {\em Phys. Rev.} {\bf
  D79} (2009) 084008, [\href{http://arxiv.org/abs/0901.3775}{{\tt
  arXiv:0901.3775}}].

\bibitem{Mohanty2011}
S.~Mohanty and S.~Rao, {\it {Constraint on super-luminal neutrinos from vacuum
  Cerenkov processes}},  \href{http://arxiv.org/abs/1111.2725}{{\tt
  arXiv:1111.2725}}.

\bibitem{Mohanty2012}
S.~Mohanty and S.~Rao, {\it {Neutrino processes with power law dispersion
  relations}},  {\em Phys. Rev.} {\bf D85} (2012) 102005,
  [\href{http://arxiv.org/abs/1112.2981}{{\tt arXiv:1112.2981}}].

\bibitem{Girelli2012}
F.~Girelli, F.~Hinterleitner, and S.~Major, {\it {Loop Quantum Gravity
  Phenomenology: Linking Loops to Observational Physics}},  {\em SIGMA} {\bf 8}
  (2012) 098, [\href{http://arxiv.org/abs/1210.1485}{{\tt arXiv:1210.1485}}].

\bibitem{Anchordoqui2014a}
L.~A. Anchordoqui, V.~Barger, H.~Goldberg, J.~G. Learned, D.~Marfatia,
  S.~Pakvasa, T.~C. Paul, and T.~J. Weiler, {\it {End of the cosmic neutrino
  energy spectrum}},  {\em Phys. Lett.} {\bf B739} (2014) 99--101,
  [\href{http://arxiv.org/abs/1404.0622}{{\tt arXiv:1404.0622}}].

\bibitem{Tomar2015}
G.~Tomar, S.~Mohanty, and S.~Pakvasa, {\it {Lorentz Invariance Violation and
  IceCube Neutrino Events}},  {\em JHEP} {\bf 11} (2015) 022,
  [\href{http://arxiv.org/abs/1507.03193}{{\tt arXiv:1507.03193}}].

\bibitem{Kostelecky2011}
V.~A. Kostelecky and N.~Russell, {\it {Data Tables for Lorentz and CPT
  Violation}},  {\em Rev. Mod. Phys.} {\bf 83} (2011) 11--31,
  [\href{http://arxiv.org/abs/0801.0287}{{\tt arXiv:0801.0287}}].

\bibitem{Gagnon2004}
O.~Gagnon and G.~D. Moore, {\it {Limits on Lorentz violation from the highest
  energy cosmic rays}},  {\em Phys. Rev.} {\bf D70} (2004) 065002,
  [\href{http://arxiv.org/abs/hep-ph/0404196}{{\tt hep-ph/0404196}}].

\bibitem{Scully2009}
S.~T. Scully and F.~W. Stecker, {\it {Lorentz Invariance Violation and the
  Observed Spectrum of Ultrahigh Energy Cosmic Rays}},  {\em Astropart. Phys.}
  {\bf 31} (2009) 220--225, [\href{http://arxiv.org/abs/0811.2230}{{\tt
  arXiv:0811.2230}}].

\bibitem{Bi2009}
X.-J. Bi, Z.~Cao, Y.~Li, and Q.~Yuan, {\it {Testing Lorentz Invariance with
  Ultra High Energy Cosmic Ray Spectrum}},  {\em Phys. Rev.} {\bf D79} (2009)
  083015, [\href{http://arxiv.org/abs/0812.0121}{{\tt arXiv:0812.0121}}].

\bibitem{Maccione2009}
L.~Maccione, A.~M. Taylor, D.~M. Mattingly, and S.~Liberati, {\it {Planck-scale
  Lorentz violation constrained by Ultra-High-Energy Cosmic Rays}},  {\em JCAP}
  {\bf 0904} (2009) 022, [\href{http://arxiv.org/abs/0902.1756}{{\tt
  arXiv:0902.1756}}].

\bibitem{Rubtsov2006}
G.~I. Rubtsov et~al., {\it {Upper limit on the ultrahigh-energy photon flux
  from agasa and yakutsk data}},  {\em Phys. Rev.} {\bf D73} (2006) 063009,
  [\href{http://arxiv.org/abs/astro-ph/0601449}{{\tt astro-ph/0601449}}].

\bibitem{Abbasi2016}
{\bf Pierre Auger, Telescope Array} Collaboration, R.~Abbasi et~al., {\it
  {Report of the Working Group on the Composition of Ultra High Energy Cosmic
  Rays}},  {\em JPS Conf. Proc.} {\bf 9} (2016) 010016,
  [\href{http://arxiv.org/abs/1503.07540}{{\tt arXiv:1503.07540}}].

\bibitem{Aab2016a}
{\bf Pierre Auger} Collaboration, A.~Aab et~al., {\it {Evidence for a mixed
  mass composition at the ‘ankle’ in the cosmic-ray spectrum}},  {\em Phys.
  Lett.} {\bf B762} (2016) 288--295,
  [\href{http://arxiv.org/abs/1609.08567}{{\tt arXiv:1609.08567}}].

\bibitem{Mattingly2008}
D.~Mattingly, {\it {Have we tested Lorentz invariance enough?}},  in {\em
  {Proceedings, Workshop on From quantum to emergent gravity: Theory and
  phenomenology (QG-Ph): Trieste, Italy, June 11-15, 2007}}, 2008.
\newblock \href{http://arxiv.org/abs/0802.1561}{{\tt arXiv:0802.1561}}.

\bibitem{Jacobson2003}
T.~Jacobson, S.~Liberati, and D.~Mattingly, {\it {Threshold effects and Planck
  scale Lorentz violation: Combined constraints from high-energy
  astrophysics}},  {\em Phys. Rev.} {\bf D67} (2003) 124011,
  [\href{http://arxiv.org/abs/hep-ph/0209264}{{\tt hep-ph/0209264}}].

\bibitem{Myers2003}
R.~C. Myers and M.~Pospelov, {\it {Ultraviolet modifications of dispersion
  relations in effective field theory}},  {\em Phys. Rev. Lett.} {\bf 90}
  (2003) 211601, [\href{http://arxiv.org/abs/hep-ph/0301124}{{\tt
  hep-ph/0301124}}].

\bibitem{Jacobson2006}
T.~Jacobson, S.~Liberati, and D.~Mattingly, {\it {Lorentz violation at high
  energy: Concepts, phenomena and astrophysical constraints}},  {\em Annals
  Phys.} {\bf 321} (2006) 150--196,
  [\href{http://arxiv.org/abs/astro-ph/0505267}{{\tt astro-ph/0505267}}].

\bibitem{Maccione2007}
L.~Maccione, S.~Liberati, A.~Celotti, and J.~G. Kirk, {\it {New constraints on
  Planck-scale Lorentz Violation in QED from the Crab Nebula}},  {\em JCAP}
  {\bf 0710} (2007) 013, [\href{http://arxiv.org/abs/0707.2673}{{\tt
  arXiv:0707.2673}}].

\bibitem{Bernstein2013}
A.~M. Bernstein and B.~R. Holstein, {\it {Neutral Pion Lifetime Measurements
  and the QCD Chiral Anomaly}},  {\em Rev. Mod. Phys.} {\bf 85} (2013) 49,
  [\href{http://arxiv.org/abs/1112.4809}{{\tt arXiv:1112.4809}}].

\bibitem{GarciaCanal2009}
C.~A. Garcia~Canal, S.~J. Sciutto, and T.~Tarutina, {\it {Testing hadronic
  interaction packages at cosmic ray energies}},  {\em Phys. Rev.} {\bf D79}
  (2009) 054006, [\href{http://arxiv.org/abs/0903.2409}{{\tt
  arXiv:0903.2409}}].

\bibitem{Letessier-Selvon2011}
A.~Letessier-Selvon and T.~Stanev, {\it {Ultrahigh Energy Cosmic Rays}},  {\em
  Rev. Mod. Phys.} {\bf 83} (2011) 907--942,
  [\href{http://arxiv.org/abs/1103.0031}{{\tt arXiv:1103.0031}}].

\end{thebibliography}\endgroup
\end{document}